\documentstyle[12pt]{article}

\evensidemargin \oddsidemargin
\catcode `\@=11
\def\numberbysection{\@addtoreset{equation}{section}
\renewcommand{\theequation}{\thesection.\arabic{equation}}}
\numberbysection
\def\be{\begin{equation}}
\def\ee{\end{equation}}
\def\bea{\begin{eqnarray}}
\def\eea{\end{eqnarray}}
\def\ll{\label}

\def\de{\delta}

\def\nn{\nonumber}

\def\ti{\tilde}

\def\ga{\gamma}

\def\ba{\begin{array}{c}}

\def\ea{\end{array}}

\def\si{\sigma}

\def\De{\Delta}

\def\ov{\over}
\def\ha{{1\over 2}}
\def\l{\left}
\def\l({\left(}
\def\r){\right)}
\def\r{\right}
\def\rw{\rightarrow}

\def\la{\lambda}
\def\al{\alpha}
\def\be{\begin{equation}}
\def\bit{\begin{itemize}}
\def\eit{\end{itemize}}
\def\ee{\end{equation}}
\def\ed{\end{document}}
\def\bea{\begin{eqnarray}}
\def\eea{\end{eqnarray}}

\begin{document}
\thispagestyle{empty}
\rightline{hep-th/9510131}
\rightline{SINP-TNP/95-15}
\smallskip

\begin{center}
{\Large \bf
Exact Bethe ansatz solution of nonultralocal quantum mKdV
model } \vspace{1.5cm} \\
{\large {\sc Anjan Kundu}
\vspace{0.3cm}\\
}
{\em
Theory Group,
Saha Institute of Nuclear Physics,\\ 1/AF  Bidhan Nagar,
Calcutta 700 064, India.}\\ {\em e-mail: anjan@saha.ernet.in}
 \end{center}
\vspace*{1cm}
\begin{abstract}
{ A lattice regularized Lax  operator  for the nonultralocal  modified
Korteweg de Vries  (mKdV) equation
is proposed at the quantum level
with the basic operators satisfying a $q$-deformed
braided algebra. Finding further the
 associated quantum $R$ and
 $Z$-matrices
the exact integrability of the model
is proved through
the braided quantum Yang--Baxter equation, a suitably
generalized equation for the nonultralocal models.
Using the algebraic Bethe ansatz
the eigenvalue problem of the quantum mKdV model  is  exactly solved and
its connection  with  the spin-$\ha$ XXZ chain is established, facilitating
the investigation of the corresponding conformal properties.
}
\end{abstract}
\setcounter{page}0

\newpage

\section {Introduction}
The theory of
 quantum  integrable systems made a major breakthrough with the formulation
 of  quantum  inverse scattering method (QISM) and the
  algebraic Bethe ansatz
 \cite{Faddeev}. These  techniques enabled us to solve exactly the eigenvalue
 problem  of a number of quantum models including the spin chains
\cite{spinchains} as well as the
field theoretic models \cite{SG}-
\cite{kul-skly}. However,  this success
apart from some recent developments was confined mostly to
 a limited class of models, known as {\em ultralocal }  models.
 For such systems
    the Lax operators at different lattice
 points $j, k ; j \neq k$  must commute, i.e.
 \be L_{bk}(\mu)L_{aj}(\la) =L_{aj}(\la)L_{bk}(\mu) , \quad a \neq b,
   \ll{ulocal}\ee
where $a,b$ signify associative space indices.
Due to this constraint  the basic equation given by the
quantum Yang-Baxter equation (QYBE)
\be
R_{ab}(\la-\mu)L_{aj}(\la)L_{bj}(\mu)=L_{bj}(\mu)L_{aj}(\la)R_{ab}(\la-\mu),
\quad
j=1, \ldots,N
\ll{rll} \ee
can be multiplied for different $j$'s and raised to its global form
\be R_{ab}(\la-\mu)T_{a}(\la)T_{b}(\mu)=T_{b}(\mu)T_{a}(\la)R_{ab}(\la-\mu)
\ll{rtt} \ee
through the monodromy matrix
\be
T_a(\la) ={L}_{aN}(\la){L}_{a,N-1}(\la)\ldots
{L}_{a1}(\la) .\ll{t}\ee
Taking trace of  (\ref{rtt}) one derives in turn  the  relation
 $[tr T(\la), tr T(\mu)]=0$,
  and since $\ln (tr T(\lambda))=\sum_{n=1}^\infty
{ C_n \ov {\la^n}}$ generates the conserved quantities,
one gets $\ [C_n,C_m]=0, \ $ for $ \ n \neq m.$ That is
the infinite number of conserved
quantities $\{C_n\}$ are in involuion, which ensures the integrability of
the corresponding quantum system represented by the Lax operator $L_i$.

However, since the above formulation of quantum integrability depends
crucially on the ultralocality condition (\ref{ulocal}),  it fails for the
{\em nonultralocal }
models like WZWN, nonlinear $\si$-model,
Korteweg de Vies (KdV), modified KdV (mKdV)
models etc. not satisfying (\ref{ulocal}). The Lax operators $L_i$  of such
 models involve
discrete fields not commuting at different lattice points, which at
the continuuum limit may correspond to  commutation relaions  like

\[ [v(x),v(y)]=\de '(x-y) \]
expressed through derivatives of the $\de$-function
\cite{Maillet}.

Nevertheless there were some progress  in treating a few
of such nonultralocal models
\cite {korepin}- \cite{goddard}
as well as  in formulating the basic theory
  \cite{Maillet91}- \cite{semenov95}, though  the
situation
   in dealing with
genuine quantum integrable models is still far from its total perfection.
 In this context  solving  quantum
KdV and mKdV models at par
with  the
elegant   approach of QISM \cite{Faddeev,SG}
remained as a major challenge.

Our aim  here  precisely is to address this problem by presenting
a quantum mKdV model and  solving its eigenvalue equation
through algebraic Bethe ansatz, much in parallel to the well established
treatment
of  sine-Gordon \cite{SG} or Liuoville \cite{fad-tim} model.
It should be mentioned  that in a very recent work an indigenous
nonultralocal formulation  of   KdV equation as a quantum mapping  was given
by Volkov \cite{volkov95}.

\section{Classical mKdV model}
Before  presenting the quantum mKdV model let us look briefly into some
of its
classical aspects as a   well known
  integrable system  \cite{Zakharov} given by the field equation
\begin{equation}
\pm v_t(t,x)+v_{xxx}(t,x)-6v^2(t,x)v_x(t,x)=0
,\ll{mkdv}\end{equation}
and  possessing an infinite set of
conserved quantities
 $C_n, \  n=1, 2, \ldots $
including the Hamiltonian
 \be H= C_3=\frac {1} {2}
 \int_{-\infty}^{\infty} dx \left( v_x^2+v^4 \right).
 \ll{H}\ee
The Poisson bracket (PB) structure for the field
 is given in a
noncanonical   form
\begin{equation}
\{ v(x),v(y) \}=\mp \delta'(x-y),
\ll{mkdvpb}\end{equation}
using which
equation  (\ref{mkdv}) can be derived from (\ref{H})
as a Hamilton equation.
Different signs in equation  (\ref{mkdv}) corresponds to fields
with different  signs in their PB relations (\ref{mkdvpb}).
Note that this noncanonical Poisson bracket (\ref{mkdvpb}) is the
source  of  the  nonultralocal  behavior  of the
mKdV model at the quantum level.

As in all   classical  integrable  systems  the
nonlinear  equation  (\ref{mkdv})  can  also  be  obtained  from the
linear system $\Phi_x=U \Phi, \ \  \Phi_t=V \Phi $ as a
flatness condition
\[U_t-V_x+[U,V]=0 \]
with the Lax pair
\be
U(v,\xi) = i\left( \begin{array}{c}
-\xi
\qquad  \ \
v \\
-v \qquad \ \
\xi
         \end{array}   \right),
\ll{U}\ee
and
\be
V(v,\xi) =\mp i\left( \begin{array}{c}
4i\xi^3 +2\xi v^2
\qquad  \ \
-4i\xi^2 v+2i\xi v_x+iv_{xx}-2iv^3 \\
4i\xi^2 v+2i\xi v_x-iv_{xx}+2iv^3  \qquad \ \
-4i\xi^3 -2\xi v^2
         \end{array}   \right).
\ll{V}\ee
It should be remembered that among this pair the Lax operator
$ U(v,\xi)$ plays more important role and  can generate alone
all  conserved quantities, which with the subsequent application of the
   PB  derives
the given equation from the Hamiltonian and    other hierachies from
the higher  conserved quantities.
The PB structure associated with the Lax operator also  carries important
information about the classical $r$-matrix.

There exists  an important connection between mKdV (\ref{mkdv}) and the
KdV equation
\begin{equation}
\pm u_t(t,x)+u_{xxx}(t,x)-6u(t,x)u_x(t,x)=0
,\ll{kdv}\end{equation}
through Miura transformation
$ \quad u=v_x+ v^2 \ \ $ which maps  the PB relation (\ref{mkdvpb})
for $ \ \  u(x)= 2 \hbar \sum_n l_n e^{-inx} \ \ $ to the classical
Virasoro algebra
\be
i \{l_n,l_m \}= (m-n) l_{n+m}+{n^3 \ov 4 \hbar} \de_{n+m,0}
\ll{virclas}\ee

\section{Lattice regularized quantum  Lax operator }

We will be interested in the quantum generalization of the mKdV model.
However our point of focus will  be not the  operator evolution equation,
but
the algebraic aspect of the problem  related to its eaxct quantum
integrability.
For this purpose
we  consider  a lattice regularized
 Lax operator
\be
L_{k}(\xi) = \left( \begin{array}{c}
(W_k^{-})^{-1}
\qquad  \ \
 i\De \xi W_k^{+}\\
-i\De \xi (W_k^{+})^{-1}
\qquad \ W_k^{-}
         \end{array}   \right),
\ll{Lmkdv}\ee
involving basic quantum  operators $W_k^{\pm},$  which   satisfy
an interesting $q$-deformed algebra
\bea
W_{j+1}^{\pm} W_{j}^{\pm} =
q^{\pm \ha}  W_{j}^{\pm} W_{j+1}^{\pm}, & &    \qquad
W_{j+1}^{\pm} W_{j}^{\mp} =
q^{\mp \ha}  W_{j}^{\mp} W_{j+1}^{\pm} \nonumber \\
W_{j}^{+} W_{j}^{-} =
q  W_{j}^{-} W_{j}^{+ }, & & \qquad
[ W_{j}^{\pm} W_{j}^{\pm} ]= 0
,\ll{W-alg} \eea
with $q=e^{ {i\hbar \ga}},$
similar to the deformed braided algebra \cite {qbraid}.
For checking that at the  continuum limit one recovers  the
 relations relevant to the mKdV  fields,  we introduce operators
$ W_{j}^{\pm}=e^{iv_{j}^{\pm} } $ with the commutators
\be
[v^\pm_k, v^\pm_l]=\mp i \ga {\hbar \ov 2} (\delta_{k-1,l}- \delta_{k,l-1})
,\ll{cr1}\ee
and \be
[v^+_k, v^-_l]=i \ga {\hbar \ov 2}
(\delta_{k-1,l}- 2\delta_{k,l}+\delta_{k,l-1})
  ,\ll{cr2}\ee
which are  consistent with algebra
(\ref{W-alg}).

At
lattice constant  $\De \rw 0,$
the discrete operators
$v^\pm_k$ go to the quantum field $\Delta v^\pm(x) $ and
due to the limit $ \partial_x {\de_{kl} \ov \De} \rw \de_x(x-y)$
the relation
(\ref{cr1})  reduces clearly to the nonultralocal commutation relation
\be
[v^\pm(x), v^\pm(y)]=\pm i \ga {\hbar \ov 2}
(\delta_{x}(x-y)- \delta_{y}(x-y)) =  \pm i \ga {\hbar } \de'(x-y)
   \ll{cr3}\ee
recovering at the classical limit   the PB relation (\ref{mkdvpb})
with $\pm$ signs corresponding to the fields   $v^\pm(x)$.
The  reduction of operators $ W_{j}^{\pm} \approx {\bf I }
+i \De v^\pm (x) $
at the continuum limit yields from (\ref{Lmkdv})
 \[ L_{k}(v^-,v^+,\xi) \rightarrow {\bf I} + \Delta {\cal L}(v^-,\xi)+
 O ({\Delta}^2 (v^+) ) \]
 with
 \be {\cal L} (v^-,\xi)=-i(v^- (x) \sigma^3+\xi \sigma^2)
\ll{lcont}\ee
linked to the well known mkdv  Lax operator  (\ref{U})
as $ U(v, \xi)=\si^1
{\cal L}(v^-,\xi)$ involving
 field $v(x),$ which is  related to  equation (\ref{mkdv}) with
$+$ and  PB (\ref{cr1}) with $-$ signs.
It is remarkable that though  the lattice Lax operator (\ref{Lmkdv})
 contains  both
$v^\pm$ variables having  nontrivial commutation relation (\ref{cr2}),
the dependence on  one of them drops out from the continuum Lax operator (in
the
present case it is $ v^+$). Moreover since (\ref{cr2}) at $ \qquad
\De \rw 0 \qquad $ yields
$ \qquad [v^+(x),v^-(y)]=  \De \cdot \partial_{xx} \de (x-y) \rw 0, $
the  fields answering to different signs of mKdV equation
and the PB relations get decouppled at the continuum limit.
Note that  interchanging the operators $W^+ \leftrightarrow W^-$
along with $q \leftrightarrow q^{-1}$, keeps the underlying algebra
(\ref{W-alg}) invariant, while  the corresponding Lax operator
is   changed as
$ L_{k}(v^-,v^+,\xi) \rightarrow  L_{k}(v^+ ,v^-,\xi)$, with the
continuum limit
\[ L_{k}(v^+,v^-,\xi) \rightarrow {\bf I} + \Delta {\cal L}(v^+,\xi)+
 O ({\Delta}^2 (v^-) ) \]
where ${\cal L}(v^+,\xi) $ dependents now on the complementary
field $v^+$ related to the $-$ sign of  (\ref{mkdv}).
We suspect that this feature is likely to be present in the lattice
regularization of all such nonultralocal nonrelativistic models
exhibiting $q$-deformed algebra.

\section{Braided QYBE and quantum  integrability of mKdV model}
\setcounter{equation}{0}
For establishing the quantum integrability of the model associated with
the Lax operator (\ref{Lmkdv}) we have to generalize the QYBE (\ref{rll})
to its braided variant \cite{kunhlav}
 \be R_{ab}(\la-\mu)Z_{ba}^{-1}L_{aj}(\la)L_{bj}(\mu)
=Z_{ab}^{-1}L_{bj}(\mu)L_{aj}(\la)R_{ab}(\la-\mu), \quad
j=1, \ldots,N
\ll{bqybel} \ee
complemented  at the same time  by the  braiding
relation
\be
L_{bj}(\mu)L_{aj+1}(\la) =L_{aj+1}(\la)
Z_{ab}^{-1}L_{bj}(\mu), \quad [L_{bj}(\mu),L_{ai}(\la)]=0, \ \ i > j+1,
\ll{braid} \ee
which  reflects
 the nonultralocal nature due to the  noncommutation  of the operators
 at  neighboring lattice points.
 In such nonultralocal models  along with the standard
$R$-matrix solution of  the Yang--Baxter equation (YBE)
\be R_{12}(\la-\mu)R_{13}(\la)R_{23}(\mu)=
 R_{23}(\mu)R_{13}(\la)R_{12}(\la-\mu)\ll{ybe} \ee
one needs to find a  $Z$-matrix
satisfying similar equations \cite{kunhlav}
\be R_{21}(\la)Z_{13}Z_{23}     =
Z_{23}Z_{13}R_{21}(\la)
\quad
 Z_{12}Z_{13}R_{32}(\la)=   R_{32}(\la)Z_{13}Z_{12}
\ll{rzz}
\end{equation}
and
\be
 Z_{12}Z_{13}Z_{23}=   Z_{23}Z_{13}Z_{12}.
\ll{zzz}
\end{equation}
In general the $Z$-matrix may also depend on the spectral parameters.
Note that the set of YBE's (\ref{ybe})-(\ref{zzz})  is compatible
with the associativity condition of (\ref{bqybel}). This formulation  is
specialized
to handle the  nonultralocality not extending beyond the next
 neighbouring points, which agrees with the present mKdV model
 with the field commutator (\ref{cr3}).

We may define  the monodromy matrix in the interval as
usual by
\[
T_a^{[k,j]}(\la) ={L}_{ak}(\la){L}_{a,k-1}(\la)\ldots
{L}_{aj}(\la)\]
acting in the space
${\cal H} \equiv V_k\otimes V_{k-1}\otimes\ldots\otimes V_j$.
It is important to note   that due to the braided QYBE (\ref{bqybel})
and  braiding relations (\ref{braid})
 the  transition to the global equation
(in analogy with the transition from
(\ref{rll}) to  (\ref{rtt})) is possible also for  this nonultralocal
case
 resulting

\begin{equation}
{R}_{12}(\la-\mu)Z_{21}^{-1} T_{1}^{[k,j]}(\la)T^{[k,j]}_{2}(\mu)
= Z_{12}^{-1} T^{[k,j]}_{2}(\mu)
T^{[k,j]}_{1}(\la){R}_{12}(\la-\mu)
\ll{rztzt}\end{equation}
where the similarity in form with  (\ref{bqybel}) reflects
 the underlying Hopf algebra structure.

For physical models with   periodic boundary condition
$ L_{aj+N}(\la)= L_{aj}(\la)$   more interesting thing happens to
 the global monodromy matrix $ T_a(\la) \equiv
  T^{[N,1]}_{a}(\la)$ for the closed chain $ [N,1] ,$ since now
 $ L_{1}(\la) $ and $ L_{N}(\la)$ becomes
nearest neighbours again with  nontrivial commutation relation  (\ref{braid}).
The corresponding global braided QYBE  is given consequently  by
\begin{equation}
{R}_{12}(\la-\mu)Z_{21}^{-1} T_{1}(\la)Z_{12}^{-1} T_{2}(\mu)
= Z_{12}^{-1} T_{2}(\mu) Z_{21}^{-1} T_{1}(\la){R}_{12}(\la-\mu)
\ll{bqybet}\end{equation}
Equipped with this theory
for quantum nonultralocal models we may start now
with the lattice Lax operator (\ref
{Lmkdv}) of the mKdV model,
which from the braiding relation (\ref{braid}) and with the
use of commutators (\ref{W-alg}) for neighboring points yields a solution
\be Z_{12}=Z_{21}=q^{-\si^3\otimes \si^3}.\ll{Z}\ee
For finding the quantum  $R$-matrix  we
 plug
  $Z$-matrix (\ref{Z}) together with $L_k$
in the braided QYBE (\ref{bqybel}) and use relation like
\[ (W^{+}_j)^{\al } (W^{-}_j)^{\beta }
 =q^{  \alpha \beta  }
(W^{-}_j)^{\beta } (W^{+}_j)^{\al}
\]
for the same lattice point along  with the trivial one
\[
 [(W^{\pm}_j)^{\al} (W^{\pm}_j)^{\beta}]=0. \]

After some algebra we arrive at the remarkable result that the
$R$-matrix of the  lattice mKdV  model is  given by the
 same  trigonometric solution

 \be
R(\lambda-\mu) = \left( \begin{array}{c}
a( \lambda-\mu)
\qquad  \ \ \qquad  \ \ \qquad  \ \    \
\\ \qquad  \    b ( \lambda-\mu)    \ \ \ c ( \la - \mu)  \qquad  \ \ \\
\qquad  \  c ( \la -\mu)  \ \ \   b ( \lambda-\mu)
\qquad \ \ \\
\qquad  \ \ \qquad  \ \ \qquad  \ \     \
a( \lambda-\mu)
         \end{array}   \right)
\ll{R}\ee
with \be
a(\la- \mu)=\sin( \lambda-\mu+\hbar\ga),\quad
b(\la- \mu)=\sin( \lambda-\mu), \quad c(\la- \mu)=\sin( \hbar\ga)
\ll{abc}\ee
as that of  the $XXZ$-spin chain.
It is an easy check  that (\ref{R}) and (\ref{Z}) thus found
satisfy all  YBE's
({\ref{ybe}-\ref{zzz}) and
therefore,  may be used
for  the braided QYBE's (\ref{bqybel}) and (\ref{bqybet})
applicable to  the periodic quantum mKdV model. Moreover
 due to the fact that  $R$-matrix
(\ref{R})  commutes with   $Z$-matrix (\ref{Z}),
the braided QYBE's in the present case
 simplify further to
\begin{equation}
{R}_{12}(\lambda-\mu) L_{1j}(\lambda)L_{2j}(\mu)
=  L_{2j}(\mu) L_{1j}(\lambda){R}_{12}(\lambda-\mu)
\ll{ybelmkdv}\end{equation}
and
\begin{equation}
{R}_{12}(\lambda-\mu) T_{1}(\lambda)Z_{12}^{-1} T_{2}(\mu)
=  T_{2}(\mu) Z_{12}^{-1} T_{1}(\lambda){R}_{12}(\lambda-\mu)
,\ll{ybetmkdv}\end{equation}
respectively.

Before  proceeding further we must justify our choice of the monodromy
matrix  taken in the same form as in the  ultralocal case:
\be
T_a(\la) ={L}_{aN}(\la) \ldots {L}_{a,k+1}(\la){L}_{a,k}(\la) \ldots
{L}_{a1}(\la).\ll{Tmkdv}\ee
Evidently
this  is possible  when   the nonultralocal property  does not spoil the
detailed structure of this global object and one gets $
  L_{k+1}L_k (\xi) \sim L_{k}L_{k+1} (\xi). $
 From (\ref{Lmkdv}) we obtain  the explicit form
 \be
L_{k+1}L_k (\xi) = \left( \begin{array}{c}
L^{(k+1,k)}_{11}
\qquad  \ \
L^{(k+1,k)}_{12}\\
L^{(k+1,k)}_{21}
\qquad \ L^{(k+1,k)}_{22}
         \end{array}   \right)
\ll{L2}\ee
where
\bea
L^{(k+1,k)}_{11}&=&(W_{k+1}^-)^{-1}(W_{k}^-)^{-1}+(\De \xi )^2
W_{k+1}^+(W_{k}^+)^{-1}
\nonumber \\
L^{(k+1,k)}_{22}&=&W_{k+1}^-W_{k}^-+(\De \xi )^2 (W_{k+1}^+)^{-1}W_{k}^+
\nonumber \\
L^{(k+1,k)}_{12}&=&i\De \xi( (W_{k+1}^-)^{-1}W_{k}^++ W_{k+1}^+W_{k}^-
\nonumber \\
L^{(k+1,k)}_{21}&=&-i\De \xi( (W_{k+1}^+)^{-1}(W_{k}^-)^{-1}+
W_{k+1}^-(W_{k}^+
)^{-1})
\ll{L2ij}\eea
 Using  nonultralocal relations (\ref{W-alg})
 one shows   that for all  expressions  (\ref{L2ij}),
  $   L^{(k+1,k)}_{ij} = q \  L^{(k,k+1)}_{ij}$  hold, which
  leads to  the
  conclusion $\   L_{k+1}L_k (\xi) = q \  L_{k}L_{k+1} (\xi) \ $
justifying the choice (\ref{Tmkdv}).

For identifying the commuting set  of conserved quantities
from (\ref{bqybet}) one may find the $k$-trace following the prescription of
\cite{skly-r, kunhlav}. However
since in our case the $R$-matrix commutes with $Z$ we may start from
(\ref{ybetmkdv}), multiply it  by $Z$ and  take the trace  as
\[ tr
(Z_{12} T_{1}(\lambda)Z_{12}^{-1} T_{2}(\mu))
= tr (Z_{12} T_{2}(\mu) Z_{12}^{-1} T_{1}(\lambda)), \]
forcing the standard relation
\[ [tr  T(\la),tr T(\mu)]= 0  \quad
  .\]
 Therefore  for
\be T(\la)=
 \left( \begin{array}{c}
A(\la)
  \ \
B(\la)\\
C(\la)
\ \ D(\la)
         \end{array}   \right),
\ll{T} \ee
$\tau (\la)= tr T(\la)=  A(\la) + D(\la) $
becomes the
 generator of the conserved quantities with
 the Hamiltonian of the mkdV system produced as  $
 \partial^{'''}_\la ln \tau (\la) \mid(\la=\infty)= C_3= H$ at the continuum
limit.
 Note  that since
 the symmetry
 of the $R$-matrix allows
 $[R_{12}, k_1k_2]=0$ with
$k= q^{-{\kappa \ov 2}\si^3},$ one may also consider
\[\tau (\la)= tr (k T(\la))= q^{-{\kappa \ov 2}}
 A(\la) +q^{{\kappa \ov 2}}
D(\la) \] also
for  generating the commuting family of conserved quantities.

\section {Algebraic Bethe ansatz solution for the eigenvalue problem}
\setcounter {equation}{0}
For  solving the  eigenvalue problem of the quantum mKdV model
\bea
\tau (\la) \mid \la_1, \la_2, \cdots,\la_m>&=&
(q^{-{\kappa \ov 2}}A (\la) +q^{{\kappa \ov 2}}
D(\la) )\mid \la_1, \la_2, \cdots,\la_m> \nn \\ &=&
\Lambda(\la)
 \mid \la_1, \la_2, \cdots,\la_m>
\ll{eigen}\eea
including that of its energy spectrum
we may proceed in a  way similar to  the standard algebraic Bethe ansatz
formalism \cite {Faddeev},
keeping at the same time  the track of the nonultralocal signatures.
Considering $B(\la)$ and $C(\la)$ as the {\em creation} and {\em anihilation}
operators of the pseudoparticles
we assume  the eigenvectors in the usual Bethe ansatz  form
\be
\mid \la_1, \la_2, \cdots,\la_m>= B(\la_1)B(\la_2) \ldots B(\la_m) \mid 0>
\ee
where $\mid 0> $ is the pseudovacuum. For calculating the eigenvalue of
$\tau(\la)=q^{-{\kappa \ov 2}} A (\la) +q^{{\kappa \ov 2}}
D(\la) $ we need to know first the commutation relations
of $A (\la)$ and $ D(\la)$ with $B(\mu), $  which may be obtained from the
braided
QYBE (\ref{ybetmkdv}) using the $R,Z$-matrices given as
(\ref{R}) and (\ref{Z}):
\bea
A (\la) B(\mu) &=&q^{-2} { a(\mu-\la) \ov b(\mu -\la )} B(\mu)A (\la)-
q^{-2}{ c(\mu-\la) \ov b(\mu-\la )} A(\mu)B (\la) \nonumber \\
D (\la) B(\mu) &=&q^{2} { a(\la-\mu) \ov b(\la -\mu  )} B(\mu)D (\la)-
q^{2}{ c(\la-\mu) \ov b(\la-\mu )} D(\mu)B (\la) \ll{ADB}
\eea
Note that the multiplicative factors   $q^{\pm 2}$ appearing
in the above expressions is the contribution
from nonultralocality. For proceeding further
along the QISM line  one must define a pseudovacuum
such that
\be T(\la)=
 \left( \begin{array}{c}
A(\la)
  \ \
B(\la)\\
C(\la)
\ \ D(\la)
         \end{array}   \right)
\mid 0 > =
 \left( \begin{array}{c}
\al^N(\la)
  \ \
*\\
0
\ \ \beta^N(\la)
         \end{array}   \right)
\mid 0 >
,    \ll{Tvac} \ee
which corresponds to
the action of $L_i$ on local $ \qquad \mid 0>_i, $ where $\qquad
\mid 0>= \prod_{i=1}^N
\mid 0>_i \qquad  $
 as
$\qquad L_i\mid 0>_i
=
 \left( \begin{array}{c}
\al(\la)
  \ \
*\\
0
\ \ \beta(\la)
         \end{array}   \right)
\mid 0 >_i.$ That is the pseudovacuum must be the eigenstate of the diagonal
elements of $L_i$ and anihilated by one of its off-diagonal elements.

However as evident from  the form of $L_k$  (\ref{Lmkdv})
naive
application of a single   Lax operator can not  construct such a
pseudovacuum. The situation is more like  the Toda chain \cite{toda}
 or the lattice
Liouville model \cite {fad-tim}. Therefore one may either follow the
functional Bethe ansatz method formulated in \cite {toda} or the remedy
suggested in \cite {fad-tim}. We adopt here the brilliant solution of
\cite{fad-tim} by considering the action of a {\em fused} Lax operator $
L^{(2)}= L_k L_{k+1}$ in constructing the pseudovacuum.

A careful analysis involving the action of the
fused Lax operator elements (\ref{L2ij})
shows that  here the pseudovacuum may be chosen in the
form
$$\mid 0>=\prod_{({k+1 \ov 2}=1)}^N \mid 0>_k$$ where
\be \mid 0>_k=\delta ( W_{k+1}^+- q W_{k}^+)
\delta ( W_{k+1}^-+ (W_{k}^-)^{-1}).
\ll{vac}\ee

For checking this claim and deriving the required relation (\ref{Tvac})
we consider the action of $L^{(k+1,k)}_{ij}$ from    (\ref{L2ij})
on    (\ref{vac}). This  gives for the diagonal elements
\bea
L^{(k+1,k)}_{11}\mid 0>_k&=&
\l((W_{k+1}^-)^{-1}(W_{k}^-)^{-1}+(\De \xi )^2 W_{k+1}^+(W_{k}^+)^{-1}\r)
\mid 0>_k
\nonumber \\
&=& -\l(1-(\De \xi )^2q\r)
\delta ( W_{k+1}^+- qW_{k}^+)
\delta ( W_{k+1}^-+ (W_{k}^-)^{-1})
\ll{avac}\eea
i.e. \be
\al(\la)= 2i e^{i(\la+ {\ga \hbar \ov 2})} \sin(\la+ {\ga \hbar \ov 2})
\ll{alpha}\ee
with $ \De \xi = e^{i \la}$ and similarly
\bea
L^{(k+1,k)}_{22}\mid 0>_k &=&
\l(W_{k+1}^-W_{k}^-+(\De \xi )^2 (W_{k+1}^+)^{-1}W_{k}^+\r)
\mid 0>
\nonumber \\
&=& -\l(1-(\De \xi )^2 q^{-1}\r)\mid 0>_k
\ll{dvac}\eea
yielding

 \be
\beta(\la)= 2i e^{i(\la- {\ga \hbar \ov 2})} \sin(\la- {\ga \hbar \ov 2}).
\ll{beta}\ee
On the other hand for the off-diagonal element we obtain
\bea
L^{(k+1,k)}_{21}\mid 0>&=&-i\De \xi \l(
 (W_{k+1}^+)^{-1}(W_{k}^-)^{-1}+
W_{k+1}^-(W_{k}^+
)^{-1}\r)\mid 0>_k
\nonumber \\
&=&
-i\De \xi \l( q^{-1}(W_{k}^+)^{-1}(W_{k}^-)^{-1}-
(W_{k}^-)^{-1}
(W_{k}^+
)^{-1} \r) \mid0>_k = 0
\ll{cvac}\eea
using the relation $(W_{k}^-)^{-1}
(W_{k}^+)^{-1}=q^{-1}
(W_{k}^+)^{-1}(W_{k}^-)^{-1},\ $
which corresponds to the anihilation
operation. The other element however gives
$L^{(k+1,k)}_{12}\mid 0>_k  \neq 0.$
The eigenvalue problem  (\ref{eigen})  of  the  mKdV
model therefore may be solved
using the
relations (\ref{ADB}) taking into account  the pseudovacuum properties
(\ref{Tvac}). Finally inserting the results   (\ref{alpha})
and (\ref{beta}) one   obtains
\bea
 \Lambda(\la)&=& q^{-{\kappa \ov 2}}\al(\la)^N \prod_j \l(q^{-2}  {a(\la_j-\la)
 \ov b(\la_j-\la)} \r)
 +
  q^{{\kappa \ov 2}} \beta(\la)^N \prod_j\l(q^2  {a(\la-\la_j)
 \ov b(\la-\la_j)} \r) \nonumber \\
 &=& 2i e^{i \la} [q^{-{\kappa \ov 2}+{N \ov 2}-2m} \sin^N(\la+ {\ga \hbar
\ov  2})\prod_{j=1}^m  {\sin  (\la-\la_j-{\ga  \hbar})  \ov \sin(\la-
\la_j)}  \nonumber \\
&+&
 q^{-(-{\kappa \ov 2}+{N \ov 2}-2m)} \sin^N(\la- {\ga \hbar
\ov  2})\prod_{j=1}^m  {\sin  (\la-\la_j+{\ga  \hbar})  \ov \sin(\la-
\la_j)} ]
\ll{spectrum} \eea
The determining equation for the rapidity parameters  $\la_j$  may be
obtained   from   the   vanishing  residue  condition of   the   eigenvalue
equation
(\ref{spectrum}) in the usual way:
\be
 q^{-\kappa+{N }-4m} {\l({ \sin(\la+ {\ga \hbar \ov  2}) \ov
\sin(\la- {\ga \hbar \ov  2})} \right)}^N=
\prod_{j \neq  k}^m \l( {\sin  (\la-\la_j+{\ga  \hbar})  \ov
\sin  (\la-\la_j-{\ga  \hbar})}\right)
\ll{bethe} \ee
Note  that  the  nonultralocal  property  of  the   model is
reflected in the $q^{-4m}$ factors appearing in the above relations.
Such  factors  were found to appear also in ultralocal models with
twisted boundary conditions \cite{twistbc,karow}. However for  the  mKdV
model   considered  here,  they depend on the total
number of lattice points $2N$ as in \cite {fad-tim}
and  more  interestingly,    on  the
number of pseudoparticle excitations $m$.
Comparison  of  (\ref{bethe}) with  the  Bethe  equation
for the   spin-$\ha$ $XXZ$
chain with twisted boundary condition
\be
q^{-\kappa}\l({ \sin(\la+ {\ga \hbar \ov  2}) \ov
\sin(\la- {\ga \hbar \ov  2})} \r)^N=
\prod_{j \neq  k}^m \l( {\sin  (\la-\la_j+{\ga  \hbar})  \ov
\sin  (\la-\la_j-{\ga  \hbar})}\r)
\ll{xbethe} \ee
reveals the startling fact that, if the extra $q$  factors  are ignored
the mKdV model on a lattice  with  $2N$    points  becomes
equivalent to the spin-$\ha$ $XXZ$ chain with total lattice points $N$.
Of  course  the hamiltonian of the respective systems corresponds
to different conserved quantities and  their expansion point of spectral
parameters
for generating conservation laws also differs. For obtaining the  mKdV
field  model  one   has to consider ultimately  the continuum
limit.

\section {Concluding remarks}

The spin-$\ha$ $XXZ$ chain is known to be connected with the
six-vertex    model, a classical statistical model in $2$-dimrnsions.
Therefore since the
finite-size corrections and the conformal properties of the six-vertex
model is
a well studied subject, the intriguing relation between
the  quantum mKdV  and the spin-$\ha$  $XXZ$  chain found here  should
be   an
important result in analyzing the  conformal  properties  of  the
mKdV model with $q=e^{i{ \pi \ov \nu+1}}.$ Let us recall that the
six-vertex model (with a seam)
yields \cite {karow} for the ground state energy
\[ E_0=Nf_{\infty}-{1 \ov N}{\pi \ov 6} c + O({1 \ov N^2}) \]
and for the excited states
\[ E_m -E_0 ={2 \pi \ov N}( \De+ \ti \De) + O({1 \ov N^2}) \qquad
 P_m -P_0 ={2 \pi \ov N}( \De- \ti \De) + O({1 \ov N^2}) \]
where $c=1-{6 \ov \nu(\nu+1)}, \ \nu=2,3, \ldots$ for $\kappa=1$
is the central charge and  $\De,  \ti
\De$  are conformal weights of  unitary minimal
models of the  corresponding conformal field theory.

The conformal properties of the mKdV model have
enhanced  importance due  to  its  mapping  into the KdV model through Miura
transformation. The KdV field in turn is  related to the generators of
the   Virasoro   algebra   answering   to   the  $2$-d  conformal
transformations \cite{Gervais, BLZ}. In this context  the continuum
limit
of  the fused Lax operator (\ref{L2}),  which
plays  a  crucial  role  in constructing the pseudovacuum and
deriving the Bethe equation  (\ref{bethe}), is also worth analyzing.
The  linear system corresponding to this Lax operator is
$\ \Phi_{k+2}(\xi)= L^{(2)}(\xi) \Phi_k(\xi)\ ,$ which
at the continuum limit would yield the relations
\be
\Phi_x(0)=-{i \ov 2 }(v^-(x) \si^3) \Phi(0), \ \quad \mbox {and} \quad
\ \Phi_{xx}(0)
=\left( \begin{array}{c}
 u(x)
\qquad  \ \
\ \ \\
\ \
\qquad \tilde u (x)
         \end{array}   \right)\Phi(0)
\ee
where $\Phi(0)=\Phi(\xi=0)$ and the fields $u, \ti u$ are
given by
\be u(x)=iv^-_x(x)+{v^{-}}^2(x), \ \ \ti u(x)=-iv^-_x(x)+{v^{-}}^2(x)
\ll{miura}\ee
i.e. they represent
KdV fields related trough
Miura transformation.
Replacement of $v^- \leftrightarrow v^+$ would generate similarly KdV fields
through $v^+$. Due to such relations between mKdV and KdV fields
and their connection with the Liouville field \cite{Gervais},
the recently discovered equivalence \cite{fad-tim} between quantum lattice
Liouville model and spin $(-\ha)$ $XXZ$ chain acquires renewed  importance.

It is  also hoped that the main results of the present work, e.g. construction
of the exact lattice Lax operator  for the quantum mKdV model,
the $R$ and $Z$ matrices as solutions of the braided QYBE, the Bethe
ansatz solution of the related
eigenvalue problem  and finally the mapping between
quantum mKdV and spin-$\ha$ $XXZ$ chain, would be of   significance
in the alternate approach of \cite {BLZ} in formulating  CFT through
quantum KdV related theory.


\newpage
\begin{thebibliography}{99}
\bibitem{Faddeev} L. D. Faddeev, Sov. Sc. Rev. C1 (1980) 107.
\bibitem{spinchains} L. A. Takhtajan and L. D. Faddeev,
 Russian Math. Survey 34 (1979)
11.

 P.P.  Kulish and E. K. Sklyanin, Phys. Lett. A70 (1979) 461

\bibitem{SG} E. K. Sklyanin,  L. A. Takhtajan and L. D. Faddeev,
Teor. Mat. Fiz.
40 (1979) 194.
\bibitem{fad-tim}  L. D. Faddeev and O. Tirkkonen, {\it Connection of
the Liouville model and the XXZ spin chain}, preprint hep-th/9506023 (1995)

\bibitem{dnls} A. Kundu and B. Basumallick,
J. Math. Phys.
34 (1993) 1252.

\bibitem{kul-skly} P. Kulish and E. K. Sklyanin,
Lect. Notes in Phys. (ed. J. Hietarinta et al, Springer,Berlin, 1982) vol. 151
p. 61.
\bibitem{Maillet} J.M. Maillet,  Phys. Lett. 162 B (1985) 137;

S. A. Tsyplyaev, Teor. Mat. Fiz. 46 (1981) 24


J.M. Maillet, Nucl. Phys. B269 (1986) 54 ; Phys. Lett. 167 B (1986) 401.

M. Semenov-Tian-Shansky Funct. Anal. Appl. 17 (1983) 259

\bibitem{korepin} V. E. Korepin, J. Sov. Math. 23 (1983) 2429.
 \bibitem{Nijhof} F.W. Nijhoff, H.W. Capel and V.G. Papageorgiou,
{\it Integrable quantum mappings },
Clarkson Univ. preprint INS 168/91 (1991)

 F.W. Nijhoff and H.W. Capel,
 {\it Integrable quantum mapping and nonultralocal Yang-Baxter structure},
Clarkson Univ. preprint INS 171/91 (1991).
\bibitem{babelon} O. Babelon and L. Bonora, Phys. Lett. 253 B (1991) 365

 O. Babelon, Comm. Math. Phys.  139 (1991) 619


 L. Bonora and V. Bonservizi, Nucl. Phys. B 390  (1993) 205.

\bibitem{alekfad} A. Alekseev, L.D. Faddeev, M. Semenov-Tian-Shansky  and
A. Volkov,{\it The unravalling of the quantum group structure in the
WZNW theory},  preprint CERN-TH-5981/91 (1991)

L. D. Faddeev,  Comm. Math. Phys.  132 (1990) 131

A. Alekseev, S. Statashvili, Comm. Math. Phys.  133 (1990) 353

B. Blok,  Phys. Lett. 233B (1989) 359



\bibitem{reshet} N. Yu. Reshetekhin and  M. Semenov-Tian-Shansky ,
 Lett. Math. Phys. 19  (1990) 133
\bibitem{goddard} M. Chu, P. Goddard, I. Halliday, D. Olive and
A. Schwimmer,
 Phys. Lett. 266 B  (1991) 71.

\bibitem{Maillet91} L. Freidel and J.M. Maillet, Phys. Lett. 262 B  (1991) 278.

 L. Freidel and J.M. Maillet, Phys. Lett. 263 B  (1991) 403

\bibitem{kunhlav} L. Hlavaty and Anjan Kundu,
{\it Quantum integrability of nonultralocal models through Baxterization
of quantized braided algebra}, preprint BONN-TH-94-15, hep-th/9406215
\bibitem{hlav} L. Hlavaty
{\it Algebraic framework for quantization of nonultralocal models},
preprint hep-th/9412142

C. Schwiebert, {\it Generalized quantum inverse scattering method }
preprint RIMS 1003, hep-th/9412237,

\bibitem{semenov95} M. Semenov Tian-Shansky and A. Sevostyanov
{\it Classical and quantum nonultralocal systems on the lattice}
preprint hep-th/ 9509026  (1995)
\bibitem{volkov95} A. Yu. Volkov
{\it Quantum lattice kdV equation} preprint hep-th/950924
\bibitem{Zakharov} S. Novikov, V. Manakov, L. Pitaevskii and V. Zakharov,
{\it Theory of Solitons } (Plenum, N.Y., 1984)

\bibitem{qbraid} S. Majid, J. Math. Phys. 32 (1991) 3246.

 L. Hlavat\'{y}, J. Math. Phys. 35 (1994) 2560.

\bibitem{skly-r} E. K. Sklyanin, J. Phys. A 21 (1988) 2375.

\bibitem{toda}E. K. Sklyanin, Lect. Notes in Phys. 226 (Springer,Berline, 1985)
p. 196
\bibitem{twistbc} A. Kl\"umper, M. Batchelor and P.  Pearce,
J. Phys. A 24 (1991) 3111;
 A. Kl\"umper, T. Wehner and J. Zittarz,
J. Phys. A 26 (1993) 2815

\bibitem{karow} A. Karowski, Nucl. Phys. B 300 [FS 22] (1988) 479

V. de Vega and  A. Karowski, Nucl. Phys. B 285 [FS 19] (1987) 619


\bibitem{Gervais} J. L. Gervais in {\it DST Workshop on
particle physics -supersrting theory} (ed. HS Mani et al, W. Sc., 1988) p. 287
; J. L. Gervais Comm. Math. Phys. 138 (1991) 301

\bibitem{BLZ} V. V. Bazhanov, S. Lukyanov and A. B. Zamolodchikov,
{\it Integrable structure of CFT, quantum KdV and therodynamic Bethe ansatz
, } hep-th/9412229.

V. A. Fateev and S. Lukyanov, Int. J. Mod. Phys. A7 (1992) 853, 1325



\end{thebibliography}
\end{document}